\documentclass[sigconf]{acmart}
\newcommand{\red}[1]{\textcolor{black}{#1}}
\newcommand{\redd}[1]{\textcolor{black}{#1}}
\AtBeginDocument{%
  \providecommand\BibTeX{{%
    \normalfont B\kern-0.5em{\scshape i\kern-0.25em b}\kern-0.8em\TeX}}}
\usepackage{csquotes,soul}
\usepackage{multirow}

\usepackage{tikz}
\usepackage{amsmath}
\usepackage{subcaption}
\usepackage{float,comment}
\usepackage{booktabs}
\usepackage{hyperref}

\usepackage{filecontents}

\newcommand{\remove}[1]{}
\newcommand{\Paragraph}[1]{{\bf \noindent #1}}

\copyrightyear{2024}
\acmYear{2024}
\setcopyright{rightsretained}
\acmConference[CHI '24]{Proceedings of the CHI Conference on Human Factors in Computing Systems}{May 11--16, 2024}{Honolulu, HI, USA}
\acmBooktitle{Proceedings of the CHI Conference on Human Factors in Computing Systems (CHI '24), May 11--16, 2024, Honolulu, HI, USA}
\acmDOI{10.1145/3613904.3642342}
\acmISBN{979-8-4007-0330-0/24/05}

\begin{document}

\date{}

\title{Understanding Underground Incentivized Review Services}

\author{Rajvardhan Oak}
\authornote{As of this writing, the author is employed by Microsoft Corporation. However, this research is not endorsed by Microsoft in any way. Opinions expressed are the author's own, and not of Microsoft Corporation.}
\email{rvoak@ucdavis.edu}
\affiliation{%
  \institution{University of California, Davis}
  \city{Davis}
  \state{California}
  \country{USA}
}

\author{Zubair Shafiq}
\email{zubair@ucdavis.edu}
\affiliation{%
  \institution{University of California, Davis}
  \city{Davis}
  \state{California}
  \country{USA}
}

\begin{abstract}

    While human factors in fraud have been studied by the HCI and security communities, most research has been directed to understanding either the victims' perspectives or prevention strategies, and not on fraudsters, their motivations and operation techniques. 
    Additionally, the focus has been on a narrow set of problems: phishing, spam and bullying. 
    In this work, we seek to understand review fraud on e-commerce platforms through an HCI lens. 
    Through surveys with real fraudsters (N=36 agents and N=38 reviewers), we uncover sophisticated recruitment, execution, and reporting mechanisms fraudsters use to scale their operation while resisting takedown attempts, including the use of AI tools like ChatGPT. 
    We find that countermeasures that crack down on communication channels through which these services operate are effective in combating incentivized reviews. 
    This research sheds light on the complex landscape of incentivized reviews, providing insights into the mechanics of underground services and their resilience to removal efforts. 

\end{abstract}


\maketitle

\section{Introduction}
\label{sec:Introduction}

While cyber security is an area of growing importance in the HCI community, most research focuses on the victims' experiences with certain kinds of cyber crime~\cite{breen2022large} or the effectiveness of intervention strategies to prevent those crimes~\cite{li2022obfu, heuer2022compa, ganesh2023tailor}. Relatively little is known about the experiences and perspectives of actors who actually perpetrate online fraud, including how they operate and more crucially, how they evade detection and combat takedown efforts. 
Additionally, most research efforts at the HCI-Security intersection have focused on a narrow set of problems: mainly phishing~\cite{distler2023theinfl, marin2023theinfl, tally2023tips} and privacy~\cite{hasan2023psycho, roemmich2023emotion}. Although more recent threats such as reputation manipulation and review fraud are growing in prevalence~\cite{jamshidi2019characterizing}, no academic work has treated this as an HCI problem, but focusing on the technical detection measures instead.

Our work aims to bridge these gaps by treating review manipulation as an HCI problem, and, in doing so, uniquely focuses on the modus operandi of the attackers involved. 
We study underground incentivized review services, which allow sellers to solicit positive reviews from real customers in exchange for free products. \redd{While there exist some forms of legitimate incentivized reviews (e.g., platform-run programs such as Amazon Vine that aim to solicit honest and unbiased opinions and are marked as ``Vine Customer Review of Free Product'' for transparency \cite{amazonvine}), our unique focus is on \textit{illegitimate} incentivized reviews that aim to solicit misleading guaranteed positive reviews and do not carry any disclosure of the free product incentive}. Since such incentivized reviews are generally prohibited on e-commerce marketplaces \cite{amazon2016policyincentivizedreviews}, buyers actually purchase the product on the e-commerce marketplace, and then get reimbursed out-of-band (e.g., via PayPal) after they provide evidence of the submitted positive review. Operating through a complex web of multiple intermediaries, a myriad of social media platforms, and spanning multiple countries, underground incentivized review services mediate the interactions between buyers and sellers while receiving a commission from the sellers in exchange for sourcing a five-star review. 

In order to understand the organization and operational characteristics of these services, the motivations and incentives for the players involved, and fraudsters' mental models on detection, we conduct qualitative surveys with review agents and reviewers who engage in incentivized review fraud. Drawing upon insights from $N=36$ review agents and $N=38$ incentivized reviewers, we discover how review rings operate and evade detection. 

Incentivized reviews have garnered significant attention over the past few years, with both the public and private sector taking steps to combat them. The Federal Trade Commission (FTC) has proposed a new set of rules~\cite{ftcanprsummary2023} through which it can prosecute businesses for failing to disclose an incentive in a review. 
Multiple e-commerce companies (Google, Yelp, TrustPilot, TripAdvisor and Amazon) responded favorably, stating that they have been removing millions of suspect reviews over the years~\cite{yelpanpr2023, googleanpr2023, tpanpr2023}. Additionally, Amazon, through a string of lawsuits sued over $10,000$ individuals who were moderators of Facebook groups involved in review brokering~\cite{amazon2022lawsuit}.
To examine the effect of the lawsuit by Amazon and the subsequent group takedown by Facebook, we conducted a follow-up survey with $N=32$ agents. Through these responses, we discovered how effective the lawsuit was, and how the fake reviews economy adapted to the changing landscape of group deletions. 

\Paragraph{Research Questions.}
Our work aims to answer the following research questions:

\begin{list}{}{}
    \item \textit{\textbf{RQ1.}} What are the demographic, behavioral and operational characteristics of the key players involved in the reviews ecosystem? 

    \item \textit{\textbf{RQ2.}} What strategies are used by agents and reviewers to evade detection of incentivized reviews?

    \item \textit{\textbf{RQ3.}} How do agents in review marketplaces adapt to the countermeasures against incentivized reviews?

    \item \textit{\textbf{RQ4.}} What is the effectiveness of existing takedown measures (both technical and non-technical) in detecting incentivized reviews?
    
\end{list}

\Paragraph{Contributions.}
Our study examines recruitment, execution and evasion strategies in underground review services.
While our investigation focuses on products on Amazon.com, evidence from our qualitative study shows that much of what we uncover also exists on other e-commerce marketplaces such as Walmart, Target, and Wayfair. Our study furthers the understanding of the inner workings of the incentivized review ecosystem. In summary, our key contributions are as follows:
\begin{itemize}
\item First, we systematically examine the inner workings of fake review rings, including the operational characteristics, interplay between the various actors involved, and their motivations for engaging in fraud. 
\item Our work discovers evasion strategies that fraudsters use to avoid detection by the e-commerce platform. We discover that fraudsters engage in purposeful manipulation of their behavior (such as tweaking their browsing activity or review content) so that incentivized reviews are not discovered.
\item We show the role that social media plays in review services, and demonstrate how certain features (such as targeted advertisements) actually enable more effective perpetration of review fraud.
%
%
\item We conduct an audit to evaluate current detection mechanisms and find that while technical countermeasures (machine learning solutions and review removal) are ineffective in countering incentivized reviews, collaborative efforts between platforms and legal or policy measures show potential for bringing review services to their decline.
\item We show how incentivized review services are adapting to the changing landscape around them (such as reacting to lawsuits by Amazon, and leveraging AI tools to enhance their fraud).

\end{itemize}

\Paragraph{Paper Organization.} 
The rest of the paper is organized as follows. 
In Section~\ref{sec:background}, we provide an overview of the reviews ecosystem, discuss prior research and the novelty of our work. 
Section~\ref{sec:Methodology} describes our study design and data collection strategies. 
Section \ref{sec:q2} draws upon our qualitative analysis to identify motivations and operational characteristics of agents and buyers.
Section~\ref{sec:q3} outlines key evasion strategies used by agents and jennies in order to avoid review detection by Amazon, followed by Section~\ref{sec:q4} that audits existing takedown measures and their effectiveness in curbing incentivized reviews.
In Section~\ref{sec:changingeco}, we present the results of our second survey and highlight approaches using which fraudsters adapt to targeted countermeasures such as lawsuits and group moderation.
Finally, we summarize our findings and conclude in Section~\ref{sec:conclusion}.

\section{Background}
\label{sec:background}


\subsection{Key Players}
There are three major players in the incentivized reviews ecosystem; sellers, agents, and jennies, which have been briefly described in prior investigations~\cite{nguyen2018inside, nguyen2019her, crockett20195star}.

\Paragraph{Sellers.} Third-party sellers on e-commerce platforms can buy products in bulk from a wholesaler like AliExpress, and use another platform (like Amazon, Target or Walmart) for sales; poor quality products can be purchased at a low price and sold at a significant margin.
%
%
Sellers stand to gain the most from positive reviews; more positive reviews usually means better ranking and the product shown to more potential buyers. 
Note that here we specifically refer to third-party vendors as sellers, and not the e-commerce platform which itself is a first-party seller (for example, Amazon acts as a platform for other sellers, but also sells its own line of products through the Amazon Basics brand). 
%

\Paragraph{Jennies.} Buyers who will buy a product and leave a five-star review for it are referred to as \textit{jennies} by the agents. 
Note that jennies are regular, paying customers of the e-commerce platform; they are sources of revenue, engagement and advertisement impressions.
A \textit{jenny} is different than a traditional crowd-worker; while crowd-workers rely on crowd-sourced tasks for their livelihood and typically do hundreds of tasks a day, jennies write the occasional fake review in exchange for a free product. 
The persistence of engagement held by a jenny is much shorter than that one would expect of a crowd-worker. Jennies engage in review fraud to get free products, and some even make a profit by reselling the products they receive~\cite{nguyen2018inside}.

\Paragraph{Agents.} These are the middle-men contracted by sellers to identify potential buyers.
Through Facebook groups, Telegram and Slack channels, Discord servers, and targeted advertisements, these agents reach out to buyers and help the sellers obtain incentivized positive reviews. 
%
%
They instruct buyers (\textit{jennies}) on how to make purchases and what evasion tactics to use to avoid detection. 
They also communicate information about orders, reviews, and refunds (which happen through a medium outside of the e-commerce platform, such as PayPal) from buyers to sellers and vice-versa.

\subsection{End-to-End Functioning}
\begin{figure}[h!]	
	\centering{
		        \includegraphics[height=8cm]{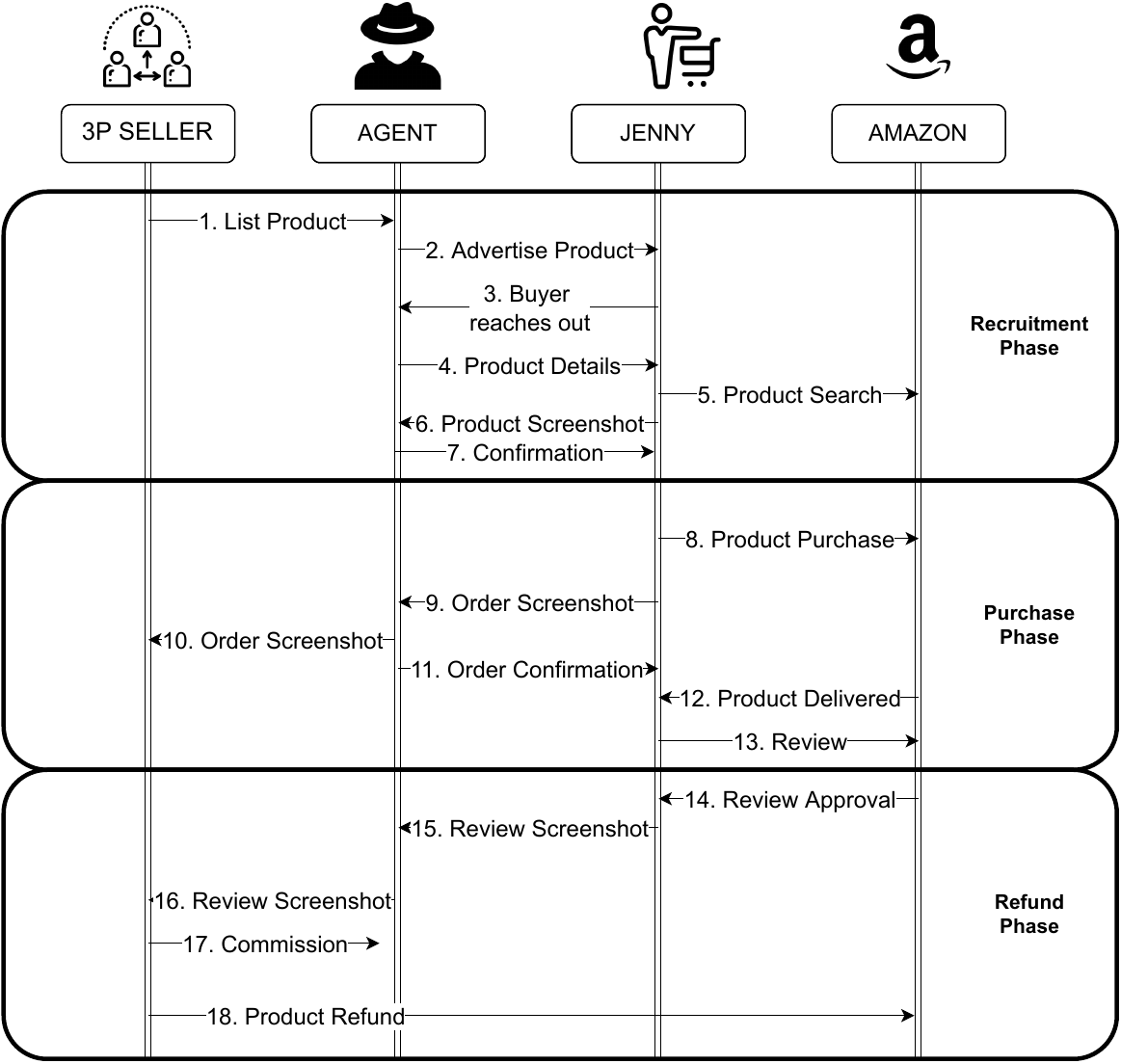}
	          } 
	\caption{End-to-End Functioning of the Reviews Market}
	\label{fig:reviews_end2end_timing}
\end{figure}

At a high level, a jenny leaves a five-star positive review for a product and gets the product for free in return. 
Figure~\ref{fig:reviews_end2end_timing} depicts how the incentivized reviews market functions end-to-end.
First, sellers identify potential buyers (also called \textit{jennies}) via their agents (steps 1 - 3). 
Agents assist jennies in searching for and buying the product (steps 4 - 8). 
The jenny then sends a screenshot of their order, which is ultimately confirmed by the seller (steps 9 - 11); the order screenshot contains an order number that the seller can use for verifying the order and tracking any return / refund activity. 
After the jenny receives the product, they submit a review. 
All reviews need to be approved by Amazon; once a review is live, the buyer sends a screenshot to the agent which is ultimately confirmed by the seller (steps 13 - 16). 
The seller now has a five-star positive review for their product; they pay the agreed-upon commission to agents and refund the buyer for the full price of the product (steps 17 - 18), effectively making it free for them; they can either use it or resell it through eBay or Facebook Marketplace and make a profit. 

\subsection{Legitimacy of Incentivized Reviews}
\subsubsection{Legitimate vs. Illegitimate Incentivized Reviews}
Incentivized reviews are a marketing strategy that is used to solve the cold start problem and encourage consumers to buy a newly introduced product~\cite{petrescu2018incentivized, marinescu2018incentives, park2020incentivized}. 
Platform-run programs, affiliate marketing, and sponsored posts on social media are all forms of incentivized reviews. 
Incentivized reviews are not inherently illegitimate. 
They are considered problematic if (a) the incentive to write the review is not disclosed or (b) the incentive depends on the sentiment of the review (i.e., the review has to be positive to earn the incentive). 
For example, in platform-run incentivized review programs such as Amazon Vine~\footnote{https://sell.amazon.com/tools/vine}, the fact that the review comes from a Vine reviewer is disclosed (``Vine Voice''), and the incentive does not depend on the sentiment of the review; Vine reviewers will get paid whether the review is positive or not.   
Our goal, however, is to study \textit{illegitimate incentivized reviews} that violate these principles. 
These incentivized reviews on Amazon are typically obtained by sellers outside the e-commerce platform's knowledge. 
Such reviews do not carry a disclosure of the incentive and are required to be five-star reviews by the seller. 
As a result, these incentivized reviews falsely appear to be organic and not driven by any incentives. 
There is evidence of the harm caused by fraudulent incentivized reviews. 
For example, an investigation by Nguyen~\cite{nguyen2018inside} revealed that a boric acid health supplement had stellar (incentivized) reviews, but was determined by doctors to be potentially fatal. 
The same investigation revealed that a local business lost its sales by more than half because of fraudulent positive reviews posted to another seller selling counterfeit goods.
\subsubsection{Legal Violations}
Major e-commerce platforms such as Walmart~\cite{walmarttos}, Amazon~\cite{amzntos}, Ebay~\cite{ebaytos} and Etsy~\cite{etsytos} explicitly prohibit such reviews in their terms of service, and such restrictions are also built into the contracts sellers have with them. 
Therefore, sellers who engage in incentivized review activity are at least in violation of the terms of their contracts (terms of service [ToS]), which itself is grounds for a civil lawsuit~\cite{amazon2022lawsuit}. 
Additionally, incentivized review activity may also amount to criminal fraud in certain jurisdictions. 
In 2022, Amazon filed a lawsuit~\cite{amazon2022lawsuit} against several incentivized review groups arguing that ToS violations may also constitute ``exceeding authorized access" that violates the Computer Fraud and Abuse Act (CFAA)~\cite{cfaa} resulting in criminal liability~\cite{cfaashifrin}. 
In June 2018, the owner of a company called PromoSalento (which offered review boosting services for TripAdvisor) was found guilty of criminal conduct on the grounds of using a fake identity to commit fraud, and sentenced to nine months in prison as a result~\cite{tpplawsuit} in the EU. 
Finally, as discussed in more detail next, incentivized review activity can be deemed ``deceptive'' and violate consumer protection laws.
For example, Amazon's lawsuit alleged violation of the Washington Consumer Protection Act because undisclosed incentivized reviews deceive consumers. 
\subsubsection{Consumer Deception}
In the United States, the Federal Trade Commission (FTC) has the authority to investigate misleading endorsements and enforce actions against businesses who engage in review fraud for deceptive practices under the Section 5 of the FTC Act~\cite{ftcact}. The FTC has exercised this power in the past~\cite{ftcpenalty} to specifically address the issue of incentivized reviews. 
Our findings have partially informed the FTC's recent publication of a new set of rules \cite{ftcanprsummary2023} that provide guidelines around incentivized reviews. 
Specifically, the FTC rules state -- among other things -- that ``it is an unfair or deceptive act or practice and a violation'' if a business does not provide ``clear and conspicuous'' disclosure that is ``easily noticeable (i.e., difficult to miss) and easily understandable'', and if a business provides ``compensation or other incentives in exchange for, or conditioned on, the writing or creation of consumer reviews expressing a particular sentiment''.
$\S$ 465.4 specifically states that ``It is an unfair or deceptive act or practice and a violation of this Rule for a business to provide compensation or other incentives in exchange for, or conditioned on, the writing or creation of consumer reviews expressing a particular sentiment, whether positive or negative, regarding the product, service, or business that is the subject of the review.''
Therefore, the lack of disclosure about the incentives and the requirement that they must be five-star constitutes a violation of the FTC regulations.

\subsection{Related Work}


\subsubsection{Fraudsters and Fraud} 
Most HCI research on cyber crime has focused on examining the prevalence of certain kinds of crime~\cite{breen2022large}, user perception of various attacks~\cite{thomas2022s, wang2022impact}, or user privacy behaviors~\cite{westin2021s, kekulluoglu2022understanding, boyd2021understanding}. However, little is known about how fraudsters organize themselves, how they function and what their motivations for fraud are. Padgett~\cite{padgett2014profiling} provides a comprehensive description of various kinds of fraud actors, their modus operandi and tell-tale signs. While the work focuses mainly on financial fraud, it describes a motivation model \red{that} profiles fraudsters as motivated by need, greed, anger or pressure. Annual reports released by the Association of Certified Fraud Examiners (ACFE) discuss characteristics of fraudsters from reported fraud, and find notably that fraudsters typically have limited education with most of them finishing only high school~\cite{bales2011evaluating}. Maimon and Louderback~\cite{maimon2019cyber} provide a detailed overview of cyber crime ecosystems, and the malicious actors involved. Financial gain and peer pressure from family were noted to be the major motivating factors for fraudsters in Nigeria, according to a study~\cite{ogunleye2019pathways} that conducted in-depth interviews with cyber criminals. Farooqi et al~\cite{farooqi2017characterizing} provide a detailed analysis of an online black-hat marketplace and study user demographics, revenue and operational characteristics for services like backlink generation, fake social media followers and spam. Along similar lines to our work, Rahman et al\cite{rahman2019art} conducted interviews with black hat app search optimization workers and discovered their operational mechanisms, mental models around detection and being flagged, as well as strategies employed in order to evade detection. 

\subsubsection{Detecting Incentivized Reviews } 

A study by Zhang et al~\cite{zhang2021detecting} collects incentivized review data from social media channels (groups on Facebook and WeChat) and by generating a set of co-review graphs, detects suspicious user communities writing reviews for a particular set of products. Similarly, He et al~\cite{he2022detecting} show that products engaging in review manipulation are highly clustered in product-reviewer networks, and therefore, network-based features can be used to detect products which have onboarded incentivized reviews. Another study~\cite{10.1145/2736277.2741650} examines underground markets for reputation escalation and finds that such services can improve seller reputation rapidly, while having a detection rate of less than $3\%$. 
%
Fake review marketplaces, especially for Amazon, have received wide media coverage and reports~\cite{nguyen2019her, crockett20195star, leticia2019some}, and reveal that Amazon is trying to crack down on these incentivized reviews using machine learning systems in combination with social media to detect them and stop them at source.
The work that is most closely aligned with ours is a 2018 investigation by Nguyen~\cite{nguyen2018inside}.
Through interviews with a few complacent buyers and analysis of social media channels for review brokering, they investigate the fake reviews market and incentives of buyers and the operation of incentivized review services through Facebook groups, Slack channels and Reddit threads. Their work shows that a family-run business for bedsheet fasteners suddenly suffered a 50\% loss in revenue because of counterfeit products; these sellers manufactured duplicate products, and obtained incentivized reviews for them which ranked them higher in a short period.

\subsubsection{Novelty and Motivation}
The prevalence of review fraud through underground services is growing day-by-day. While media investigations have discovered some characteristics for reviewers, our work is the first to systematically examine the reviews ecosystem, and gather insights from both agents and jennies. Academic research on fraudsters has been focused on phishing, romance or financial scams and never on fake reviewers. As a result, our understanding of the underground reviews economy is severely limited and consists only of anecdotal reports. Additionally, as we note above, fraudsters under consideration are not typical fraudsters; they are actual users of Amazon.com, with some fraud activity interleaved into legitimate activity on their accounts. This makes them particularly interesting to study as they are likely to exhibit characteristics and motivations different than traditional fraudsters. To the best of our knowledge, no prior work has sought to understand review services and discover motivations of such fraudsters and their evasion strategies.

\section{Methodology}
\label{sec:Methodology}
In this work, we study the functioning of the fraudulent reviews ecosystem. We use Amazon as a case study because \textit{(i)} It is the largest e-commerce platform in the US, and \textit{(ii)} it appeared to be the most favoured by sellers and agents in seeking incentivized reviews. In order to closely understand the inner-working of the incentivized review ecosystem, and examine the motivations and key tactics involved, we conducted semi-structured (consisting of both multiple-choice and free form response questions) surveys with agents and jennies.
%

\subsection{Data Crawling}

We identified groups on Facebook where agents offer products for review (more information about groups in Section~\ref{sec:targeting_jennies}). 
We ordered the groups by the number of members and average daily posts and chose the ones which were reasonably active. 
Based on the distribution in members and daily active posts, we defined a group to be active if it had more than $1500$ members and at least $10$ posts a day. 
This resulted in $156$ groups. 
\red{
We posed as buyers from the United States and observed group activity to identify who the agents and buyers were. The groups serve as a hub for agents and buyers to interact with each other. Agents frequently advertise the products listed, and share links to spreadsheets that contain the list of products that need a five-star review. An example of this can be seen in Fig~\ref{fig:agentlinks} in Appendix~\ref{app:datacrawling}. Through agent posts and comments, we were able to obtain $8$ spreadsheets through links that were shared with public read permissions. In all, we were able to compile a list of $1600$ unique products. 
}
We then systematically crawled product metadata (e.g., price, seller, variants) as well as reviews for each product including the review text, posted date, additional media attachments, and helpful votes. \red{Our crawl was conducted during February--April 2022}. 
Because this list of products was obtained directly through agents, we have high confidence that the reviews of these products have been manipulated through incentivized review services. 


Six of links we obtained were links to sheets that contained multiple tabs, and information other than product details. 
These additional tabs contained links to other folders and documents as well. 
We collected all of this information manually and downloaded all files we could find. 
This supplementary material we crawled includes: (i) slide decks and video tutorials on how agents should communicate with jennies and form a network of jennies, (ii) what instructions are to be followed so that Amazon does not delete the review, and (iii) agent leader boards containing information about how many reviews each agent obtained and the bonus they earned. 
In our qualitative analysis, we draw on these sources as materials complementary to our survey responses.

\subsection{Participant Recruitment}
We recruited participants for our study using Facebook groups where reviews are brokered by agents. We used the search term \textit{"Free Amazon Products"} and ordered the search results by number of group members and post frequency. We joined the top ranking group ($900k$ members and $78$ posts per day on average) as it would provide a large sample to study the fake reviews market. 
By observing the posts and group activity, we identified which group members were agents (members who posted pictures of products and asked for reviews) and which of them were jennies (members who expressed their interests in receiving those products through comments on the posts). 
For every potential participant, we examined group activity over the past two weeks, and selected those which were fairly active (at least $10$ posts or comments on $10$ unique posts in the two-week period). 
Following this criteria, we identified the top $500$ agents and jennies, and reached out to these individuals via Facebook Messenger. 
Potential participants were provided the details on the nature of our study and assured that we would not collect any PII (name, email, IP address). 
Once they consented, we sent them a link to our survey. 
All participants were over $21$ years of age (self-reported).

\subsection{Survey Design} 
Our surveys were conducted online via Google Forms in two phases.
In the first survey conducted in March 2022, we posed questions in four main areas: operations (how the process works, key players), demographics (age, location, gender identity), incentives (motivations and earnings), and experience and evasion (tactics to avoid detection, observations, mental models). The goal of this survey was to characterize the key players in underground review services and their manipulation strategies.  
In July 2022, Amazon filed a lawsuit~\cite{amazon2022lawsuit} against more than $10000$ agents, who were moderators of Facebook groups. According to the lawsuit, Amazon investigators worked with Facebook to have those groups deleted. To understand the perceptions, impact and response of the services around these targeted takedowns by Amazon, we conducted a follow-up survey with agents in January 2023.
We surveyed a total of $N = 36$ jennies and $N = 38$ agents in the first survey. We surveyed $N = 34$ \red{agents} in the second survey; however, as we did not collect any personally identifying information, we were unable to link these responses to the first survey. We refer to the agents and jennies in the first survey as participants \textit{A1 - A38} and \textit{J1 -- J36} respectively. We refer to agents in the second survey as \textit{B1 -- B34} to distinguish them from the agents in the first round.
%
%
%
All of the questions asked are described in Appendix~\ref{app:survey_questions}. 

\subsection{Analysis}
The surveys were semi-structured, meaning that some responses were free-form. 
In order to identify key themes in our free-text survey responses, we used iterative coding inspired by grounded theory~\cite{chun2019grounded}. Two coders were involved in the analysis.
Initially, the first coder analyzed our responses in batches of $5$, and identified codes for every free-text response. 
In every batch, they recorded earlier batches if new codes had been found. The second coder then used the code book created and performed deductive coding independently. The inter-coder agreement was high ($>0.95$), indicating \red{ a substantial strength of agreement between both coders and the reliability of our coding.}.
We identified codes in broadly three phases of the incentivized reviews process: recruitment phase, purchase phase and review phase and in roughly three themes: onboarding, motivation and evasion. 
The code book can be found in the Appendix~\ref{app:codes}.

\subsection{Ethical Considerations} 
Our recruitment strategy, surveys, and overall study protocol was reviewed and approved by the Institutional Review Board (IRB) at our institution. 

We did not collect any personally identifying information during the course of this study, and as a result, we are able to maintain participant anonymity. All the participants were aware that the answers they provide will be reported as part of a research study. Participants were recruited and shown the survey only after they provided us consent. All participants had the choice to decline to answer any of the questions, or withdraw their participation at any time during the survey. We treated all participants and their responses with respect, and followed principles laid down by the Menlo Report~\cite{kenneally2012menlo}. 

\redd{
We analyzed the content posted in various Facebook groups to curate our list of fraudulent products. These groups were public and open to join for any individual with a Facebook account. Agents marketed their products through public posts and comments in the group, which could be read by all members (an example is shown in Figure~\ref{fig:agentlinks}). Therefore, we did not engage in deception to curate out list of fraudulent products or conduct our analysis. }

Finally, throughout the course of our study, we identified several sellers and brands which engage in buying reviews for their products sold on Amazon. We do not reveal who these sellers are; if we were to do so, the sellers could trace the source back to our participants, which would jeopardize their anonymity. However, in order to further research in detecting incentivized reviews, we will share de-identified datasets, interview transcripts, key operational features and evasion tactics. Additionally, we were invited by Amazon to discuss our research with the review integrity team, where we shared insights from our work and disclosed some key discoveries.

\section{Operational Characteristics of Review Services}
\label{sec:q2}
In this section we examine the characteristics of the key players involved in underground review services, such as their demographics, operational characteristics, recruitment strategies and motivations.

\subsection{Demographics}
\label{sec:demo}
\begin{table*}[]
\begin{tabular}{|l|l|l|l|}
\hline
\multicolumn{1}{|c|}{\textbf{Attribute}} & \multicolumn{1}{c|}{\textbf{Response}} & \multicolumn{1}{c|}{\textbf{Jennies}} & \multicolumn{1}{c|}{\textbf{Agents}} \\ \hline
                  &                   &             &             \\
\textbf{Gender}   & Male              & 14(38.89\%) & 13(33.33\%) \\
                  & Female            & 19(52.78\%) & 26(66.67\%) \\
                  & Prefer not to say & 3(8.33\%)   & 0(0\%)      \\ \hline
\textbf{Age}      & 18 - 21           & 0(0\%)      & 3(7.69\%)   \\
                  & 22 - 32           & 21(58.33\%) & 31(79.49\%) \\
                  & 33 - 43           & 12(33.33\%) & 1(2.56\%)   \\
                  & 44 - 54           & 3(8.33\%)   & 3(7.69\%)   \\ \hline
\textbf{Education Level}                 & High School or Lower                   & 0(0\%)                                & 19(48.72\%)                          \\
                  & Diploma           & 0(0\%)      & 20(51.28\%) \\
                  & Bachelors Degree  & 15(41.67\%) & 0(0\%)      \\
                  & Masters Degree    & 16(44.44\%) & 0(0\%)      \\
                  & Doctorate Degree  & 5(13.89\%)  & 0(0\%)      \\ \hline
\textbf{Location} & Bangladesh        & 0(0\%)      & 18(46.15\%) \\
                  & Pakistan          & 0(0\%)      & 21(53.85\%) \\
                  & United States     & 24(66.67\%) & 0(0\%)      \\
                  & Canada            & 7(19.44\%)  & 0(0\%)      \\
                  & United Kingdom    & 5(13.89\%)  & 0(0\%)    \\ \hline 
\textbf{Annual Income (USD)} & \$0 & 0(0\%)                           & 8(20.51\%)                       \\
                                   & $1 - $1000                             & 0(0\%)                           & 10(25.64\%)                      \\
                                   & $1001 - $2000                          & 0(0\%)                           & 16(41.03\%)                      \\
                                   & $2001 - $3000                          & 0(0\%)                           & 4(10.26\%)                       \\
                                   & $3001 - $4000                          & 0(0\%)                           & 2(5.13\%)                        \\
                                   & $4001 - $50,000                        & 3(8.33\%)                        & 0(0\%)                           \\
                                   & $50,001 - $100,000                     & 8(22.22\%)                       & 0(0\%)                           \\
                                   & $100,001 - $150,000                    & 7(19.44\%)                       & 0(0\%)                           \\
                                   & $150,001 - $200,000                    & 6(16.67\%)                       & 0(0\%)                           \\
                                   & $200, 001 - $250,000                   & 4(11.11\%)                       & 0(0\%)   \\ \hline
\end{tabular}
\caption{Demographic Characteristics of Agents and Jennies}
\label{tab:demographics}
\end{table*}

\Paragraph{Sellers.} 
All of the agents we surveyed reported that the sellers they work with are based in China. 
The focus of this work is incentivized reviews on Amazon; however the underground reviews economy spans other platforms like Walmart (\textit{A8, A12, A35, J36}), Wayfair (\textit{A20, J1, J19, J33}) and Target (\textit{A3, A25, J14, J32}). 
$9$ agents reported that a seller often lists the same product on multiple e-commerce platforms.

\Paragraph{Agents.} The agents we surveyed are based in Pakistan ($54\%)$ and Bangladesh ($46\%)$. %
%
Agents are $21-35$ years old ($\mu = 25.6$). 
Agents work in groups and each group has an \textit{admin}. 
The admin is the interface between agents and the seller. 
An admin and his group of agents works for multiple sellers at the same time. 
%
%
Agents tend to have limited education; $20$ agents had earned a diploma (equivalent to an associate degree in the U.S.) and $19$ agents had only a high-school level education. 
Only $5$ agents were pursuing further education (a diploma or bachelors degree). Detailed agent demographics can be seen in Table~\ref{tab:demographics}.

\Paragraph{Jennies.} Jennies we surveyed were Amazon account holders based in the US, UK, and Canada.
However, agents reported that they also target jennies in Ireland, France, and Germany as well since some products are geared towards those countries. 
%
%
Jennies were $22-51$ years old ($\mu = 31.6$). All of the jennies held at least a bachelors degree, with $15$ holding a masters degree and $5$ holding a doctorate. 
%
This level of education is significantly higher from that observed among fraudulent actors in prior work~\cite{bales2011evaluating, padgett2014profiling}. 
With their unusually high educational qualifications, it is likely that the jennies intricately understand the nature of fraud they are perpetrating, the safeguards that might be in place to detect it, and purposely manipulate their activities to evade them. In fact, when asked about how their education shapes their understanding of review deletion by Amazon, one jenny (\textit{J15}) who holds a bachelors degree in a technical field, says:
\begin{center}
    \textit{"...I studied machine learning and web dev in my college....I built a classifier for fake review detection for my bachelors thesis project. So very helpful in knowing what to avoid..."}
\end{center}
Detailed jenny demographics can be seen in Table~\ref{tab:demographics}.


\subsection{Recruitment}
\label{sec:targeting_jennies}
Agents are responsible for recruiting jennies and help them buy products. 
There are two mediums by which agents recruit reviewers: social media channels and targeted advertisements.

\Paragraph{Facebook Groups.} 
%
We were able to find $156$ active Facebook groups which served as a hub for reviewers and sellers to contact each other. The largest of these groups had close to one hundred thousand members. 
%
On these groups, agents post information about the products which they have. Typically, agents post products on the group along with what the seller is looking for; it can be a five-star review, a five-star rating, positive seller feedback, upvotes to existing reviews or answering questions in the Q\&A section (\textit{J24, J25, J28, J30, J33}). 
Additionally, reviewers can also ask for any products they want. For example, a reviewer can make a post sharing that they are an Amazon account holder in the U.S. and are looking for bluetooth speakers. Any agents who \red{h}ave the product can comment or reach out. Jenny \textit{J30} says:
\begin{center}
    \textit{"...wanted to set up my home office, and \red{h}ad written a few reviews before. So I posted as\red{k}ing for a standing desk and office chairs - my inbox was flooded with agents who had these products..."}
\end{center}
Agents are aware that what their activities are potentially illegal, and hence, always operate with an alias. 
They also use minor perturbations on the keywords so that they may not be flagged by Facebook (see Fig.~\ref{fig:fb_group_2} in Appendix~\ref{app:recruit}).
Jennies were part of at least $1$ and at most $25$ such groups ($\mu = 6$) on Facebook.
Some jennies also reported that both jennies and agents post about scammers \red{in the groups} and warn others not to trust them (\textit{J4, J9, J21, J26, A11, A12, A29, A31}). Jenny \textit{J4} notes:
\begin{center}
    \textit{"An agent offered me a very lucrative deal and then disappeared after I sent them the review screenshot. So I immediately posted his profile screenshot on the group. Many fellow reviewers commented saying that they were talking to the same agent, and will now be cautious..."}
\end{center}
Agents also use groups to report fraudulent jennies; agent \textit{A29} says:
\begin{center}
    \textit{"If some jenny returns the product after refund, we lose commission....so I post on the group, and also include their PayPal (so they cannot just change profile name)..."}
\end{center}

The goal of the groups is to function as an exchange where reviews are bought and sold (\textit{J1, J11, J13, J15}). Groups are the most important means for agents to continue their operations and keep marketing new products to jennies which they have already identified. Through auxiliary materials shared by agents, we discovered extensive training material complete with the scripts and scenarios on how agents should operate on groups, how they can identify and approach a jenny, and how to walk them through the process.

\Paragraph{Targeted Advertisements.} Most jennies ($75\%)$ reported that they were introduced to the incentivized reviews economy via targeted advertisements on Facebook. 
Facebook facilitates the delivery of advertisements to a targeted audience; therefore agents are able to leverage search history, marketplace activity and other data available to Facebook to identify users who might be interested in writing such incentivized reviews. 
Further, a direct fallout of the targeted advertisements is the \textit{rabbit hole} phenomenon; once a user clicks or interacts positively with an advertisement for free products, they go down the rabbit hole and see several more advertisements of similar pages. As \textit{J1} says:

\begin{center}
    \textit{"I just chanced upon this side-hustle one day; I clicked on the advertisement, and then saw a few more. Pretty soon, my feed was full of such advertisements; every third or fourth post I saw was an ad for free products in exchange for reviews."}
\end{center}

As Instagram is also owned by Facebook, jennies reported that once they interacted with a few advertisements on Facebook, they started getting exposed to similar ads (advertising free products in exchange for reviews) on Instagram as well (\textit{J13, J16, J22}). 
An example of this is shown in Figure~\ref{fig:fb_ad_2} in the Appendix~\ref{app:recruit}.

\subsection{Motivations \& Incentives}
\subsubsection{Agent Incentives}
Incentives for agents are mainly monetary. Agents we surveyed are located primarily in low-income countries. Agents earn either $\$4$ or $\$5$ as commission on a review they helped procure, with the average being $\$4.42$. 
The annual income of agents (apart from the reviews) ranges from $\$0$ to $\$3600$ ($\mu$ = $\$1203$). 
Most agents earn commission comparable to their monthly income; income from reviews ranges from $33\%$ to $100\%$ of their total income. 
$7$ of the agents have no other full-time job; the reviews are their only source of income. All of the agents were aware that by facilitating incentivized reviews, they were violating Amazon's terms of service. There are three important factors that contribute to agents' motivation: additional income that is comparable to full-time salary (15 agents), lack of requirement of any infrastructure (9 agents), and lack of special education or skills (11 agents). According to agent \textit{A6}, based in Bangladesh:
\begin{center}
    \textit{"One month I earned almost \$250; my regular job pays me around \$100 a month. And it requires no additional expense; just an internet connection."}
\end{center}
Additionally, through the proxy materials, we discovered that there is an incentive structure to encourage agents to seek more reviews. Leader boards updated in real-time list agent names and number of reviews they helped procure. The top $3--5$ agents at the end of every month receive additional commission (\textit{A13, A19, A34}).

\subsubsection{Jenny Incentives}
Major motivations for jennies were the opportunity to get products for free, and not having to pay for some expensive products (such as robot vacuum cleaners (\textit{J7}), treadmill (\textit{J14}) and blenders (\textit{J17})). \textit{J8} says:
\begin{center}
    \textit{"I know it's wrong....but when I moved into my new place, I was able to get a chair, desk, humidifier, kitchen utensils, storage boxes, vacuum cleaners and a lot more for free. I saved so much by just writing reviews."}
\end{center}
Other buyers go a step ahead. They get their products for free by writing reviews, and then sell them online. Jennies reported selling products on sites like eBay, Facebook Marketplace and Offerup. Jenny \textit{J2} tells us:
\begin{center}
    \textit{"The product is basically free. Once I get my refund, I sell it on Facebook marketplace for around $75\%$ of the price. I made over \$300 last month."}
\end{center}

When asked if they were aware that writing incentivized reviews was against Amazon's terms of service, $32$ jennies ($>86\%$) responded \textit{Yes}. $4$ responded as \textit{Maybe} and only $1$ said \textit{No}. 

We also see that the financial incentives for jennies are \textit{weaker} than they are for our agents. 
Jenny incomes range from $\$25,000-\$230,000$ annually. Our lowest-earning jenny who is a graduate student earns $\$25,000$ a year, and the amount they save by writing reviews ($\$1200$ in a year) is less than $5\%$ of their annual income. 
This is in stark contrast with agents where the money earned via reviews forms at least a third of their total income (and note that agents earn a smaller amount per review than jennies save!). 
\section{Evasion Tactics}
\label{sec:q3}
In this section, we discuss the evasion tactics that jennies are asked to follow. Agents and sellers provide these guidelines during the purchase and before writing a review. The goal is to minimize the likelihood of the purchase and the following review being detected and subsequently deleted by Amazon.
By analyzing our free-text survey responses, we identified the following evasion tactics.

\begin{table*}[]
\centering
\begin{tabular}{|c|c|c|}
\hline
\textbf{Tactic} & \textbf{\textit{n} (Agents)} & \textbf{\textit{n} (Jennies)} \\ \hline
Search product with keywords &11 ($\textit{28.94\%}$) &10 ($27.78\%$) \\ \hline
Spend time reading features and reviews & 7 ($18.42\%$) & 7 ($19.44\%$) \\ \hline
Mark reviews as helpful and ask questions in the QnA &12 ($31.57\%$) &8($22.22\%)$ \\ \hline
Add similar products to saved lists and shopping cart &8 ($21.05\%$) &9 ($25\%)$ \\ \hline
Not paying with gift cards or coupons &6 ($15.79\%$) &2 ($5.5\%$) \\ \hline
Wait $10-15$ days to submit a review &11 ($28.94\%$) & 4 ($11\%$) \\ \hline
Add photos and videos to submitted review &13 ($34.21\%$) &12 ($33.33\%$) \\ \hline
Write reviews of at least $300$ words &8 ($21.05\%$) &1 ($2.25\%$) \\ \hline
Avoid buying from the same seller &3 ($7.89\%$) &2 ($5.5\%$) \\ \hline
Write some 1-star and 2-star reviews for other products &13 ($34.21\%$)  &11($30.55\%$) \\ \hline
Be consistent across all reviews in terms of timing, content, media &0 ($0\%$) &6($16.66\%$) \\ \hline
\end{tabular}
\caption{Key Evasion Tactics employed by Jennies and recommended by Agents to avoid detection of reviews by Amazon. }
\label{tab:evasion-tactics-codes}
\end{table*}

\subsection{Organic Search for Product}
Agents recommend buyers to search for the product with keywords, rather than provide a direct link or the brand name. Buyers are then asked to scroll through the products and find the correct item. According to agents, this is because Amazon can track that you landed on that product from a link, and can consider it to be suspicious. An agent (\textit{A3}) notes:
\begin{center}
    \textit{"It is for security of your account. Sellers don't allow us to share direct links with buyers, as Amazon will detect this, and may remove the review that you write."}
\end{center}
Similarly, buyers are advised never to search for a product with the brand name in the search term. As far as possible, agents never directly disclose the brand name. A jenny(\textit{J2}) describes this:
\begin{center}
    \textit{"They never show us the brand. In the image they send us, the brand name is blurred out. I try to find the correct product and send them a screenshot or link to confirm."}
\end{center}

\subsection{Platform Engagement}
As an extension of above, jennies are encouraged to engage with the Amazon platform while searching for the product as a legitimate buyer would. They are asked to look at similar products, browse through images, read reviews and simulate an organic product discovery experience. \textit{A2} says: 
\begin{center}
    \textit{"We always ask buyers to spend at least 1 minute browsing similar products, and at least 15 seconds on checkout page since if search and purchase is done too quickly, it is suspicious to Amazon."}
\end{center}
Jennies generally follow the guidelines, since their accounts are at risk. A jenny (\textit{J18}) also concurs, saying:
\begin{center}
    \textit{"I spend a lot of time in searching for the products. I really read 5-10 5* and some 1* reviews. I also ask a question or two in the forum. For some of my trusted agents, I also write reviews AFTER the return period. That will tell Amazon that my product is not probably refunded (if it was, I wouldn't risk losing the 30 day return period)."}
\end{center}
 
\subsection{Payment Restrictions}
\label{sec:paymentrestrictions}
Jennies are advised not to pay using gift cards, and always make the full payment using a different payment method. In addition, Amazon at times displays coupons using which buyers can get a discount on the product; buyers are cautioned not to use these either. The reason behind this is that the seller has designated a fixed number of coupons, and these are to attract other, genuine customers to buy the product. An agent (\textit{A7}) says:
\begin{center}
    \textit{"Most sellers tell us not to ask buyers to use coupons. If they do, they still refund them, as it is not a loss. But the coupons are mainly there for other customers (who we will not be refunding) to think that they are getting a discount."}
\end{center}
Our hypothesis behind sellers not allowing the use of a gift card is that since a gift card works via a coupon code, it could be construed by Amazon as a discount offered by the seller. 

\subsection{Review Timing}
Jennies are recommended to wait at least $10$ days before submitting a review on Amazon. Reviews which are submitted too soon arouse suspicion. According an agent (\textit{A3}):
\begin{center}
    \textit{"...Please submit reviews 10-15 days after you receive the shipment. Use the product for a few days and then review. Else, your review will be removed by Amazon."}
\end{center}
Buyers are, of course, eager to receive the refund as soon as possible. However, they understand the reasoning. Jenny \textit{J7} says:
\begin{center}
    \textit{"It makes sense. I need to have experienced the product before writing a review, otherwise it is suspicious. I maintain a spreadsheet of items I purchased along with their dates, so I know when 10 days have passed to write the review."}
\end{center}

\subsection{Review Content}
\label{sec:reviewcontent}
Agents provide detailed instructions to buyers on the content of the reviews. According to agents, factors like the length of the review, and presence of supplementary media (images and videos) affects whether Amazon will remove the review. All ratings have to be five-star. Agents always recommend writing a review of at least $300$ words. A jenny (\textit{J6}) tells us:
\begin{center}
    \textit{"Picture and video reviews are really important to them. Once, an agent offered me additional commission of \$5 if I wrote a review of more than 200 words. And \$5 more if I attached a picture or video to it."}
\end{center}
However, $5$ jennies reported that they do not follow this advice because they find it challenging to come up with a long, positive, five-star fake review (\textit{J1, J8, J12, J25}).
In addition to review length, they are cautioned never to strongly emphasize any particular seller. An agent (\textit{A11}) notes:
\begin{center}
    \textit{"Never ever mention ''always buy from this seller'' in your review. We want the product from our seller to be highly ranked. If you write about a seller, Amazon moves the review to seller feedback, which sellers don't want, because customers don't view seller feedback often."}
\end{center}

\subsection{Seller Diversity}
In general, buyers are not permitted to review multiple products from the same seller within a certain period of time. According to agent \textit{A8}:
\begin{center}
    \textit{"Very suspicious if you buy too many products from same store. I never return to the same buyer before 2 months of the first review."}
\end{center}
Buyers have another thumb rule which they follow, and that is avoiding buying too many products offered by the same agent. Jenny \textit{J10} shares their experience:
\begin{center}
    \textit{"I purchased a vacuum cleaner and sent the order screenshot to the agent. She confirmed my order, and showed me some more products to buy. When I went to amazon, I saw the SAME product as a suggestion, under the heading ''people with similar purchases to you also bought....''. It would have been very suspicious. Clearly, a pattern had been captured."}
\end{center}

\subsection{Account Metadata}
Agents and sellers also take into consideration account metadata while analyzing whether a particular buyer's review is likely to be removed or not. Of our $38$ participating agents, $31$ reported that they strongly prefer Amazon Prime account holders. According to agents, reviews from accounts who have Amazon Prime subscriptions are less likely to get removed. $24$ of our participating agents reported that when some products had some reviews removed, all of them were from non-Prime accounts.  
In addition, agents also check if a prospective buyer is an \textit{easy rater}, that is, they have too many five-star reviews and no three or four-star reviews. 
Agents also recommend that reviewers write reviews for items that they have purchased organically (that is, without rebate). They also recommend that not all reviews be five-star. Jennies are aware that a diversity in ratings is important to keep up the impression that reviews are genuine. Jenny \textit{J16} says:
\begin{center}
    \textit{"I have some non-5* reviews on my profile. If I write only 5*, amazon will think I am writing fake reviews, if I add some 1,2* they will think I am doing it to balance out, but I write some 3/4*; there is no clear incentive for me, which increases their confidence that I am a genuine buyer."}
\end{center}
We would like to note here how this need for rating diversity can be harmful for other sellers. $11$ jennies reported that for the products which they purchased by themselves, or were not incentivized, they left a negative review \textit{even if the product was good} so that they would have some 1-star and 2-star reviews on their reviewer profile. 

\subsection{Review Consistency}
This strategy is unique because none of the agents expressed it as an evasion tactic, but $8$ jennies shared that they kept all of their reviews consistent in terms of content, length, tone, and media attachments. Jenny \textit{J34} says:
\begin{center}
    \textit{"I try to be consistent across all reviews so amazon doesn't feel I'm giving special attention to any product. All my reviews are same length, all have photos, etc."}
\end{center}
Having these attributes consistent over various products (both incentivized and genuine purchases) makes it harder for Amazon to detect the incentivized reviews as different. It is interesting to note that jennies are able to understand the significance of consistency and employ it as an adversarial tactic.
\section {Effectiveness of Takedown Measures}
\label{sec:q4}

\subsection{Incentivized Review Detection by Amazon}
\label{sec:auditamz}

\begin{figure}[!h]	
	\centering{
		        \includegraphics[height=6cm]{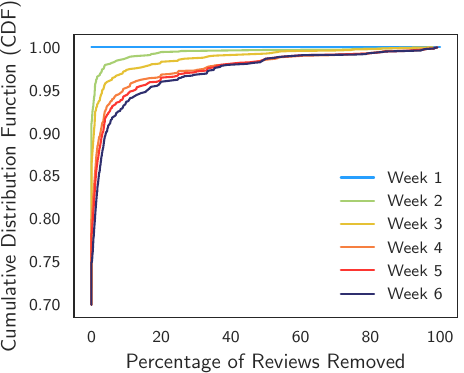}
	          } 
	\caption{Week-over-Week Deletion of Reviews}
	\label{fig:wow_deletion_curve}
\end{figure}

In this section, we evaluate how effective Amazon is at detecting incentivized reviews. 
To this end, we periodically scraped product reviews for all products in our incentivized products dataset every week over a $6$-week period. 
This allowed us to examine how many reviews were added and deleted every week. 
%
While we cannot evaluate how \textit{precise} their removal was (since we do not have access to \textit{all} reviews they removed), we can evaluate the \textit{recall} on our dataset of incentivized products. 
Figure~\ref{fig:wow_deletion_curve} shows how reviews were deleted over the 6-week period. 
We considered all of the reviews in our original crawl as a baseline, and computed the proportion of reviews removed relative to this baseline during each week. 
We then plot an empirical cumulative distribution function over products for every week. 
We can see that nearly $50\%$ of the products seeking incentivized reviews had none of their reviews removed. 

\remove{
\begin{figure}[!h]	
	\centering{
		        \includegraphics[height=5cm]{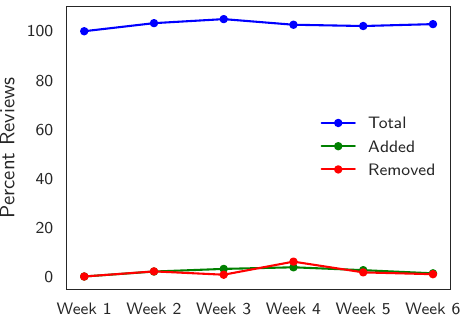}
	          } 
	\caption{Week-over-Week Review Trends }
	\label{fig:wow_addition}
\end{figure}
}

\subsection{Temporal Analysis}
\begin{figure*}[!ht]
    \centering
    \begin{subfigure}[b]{0.3\textwidth}
                 \centering
                 \includegraphics[width=\textwidth]{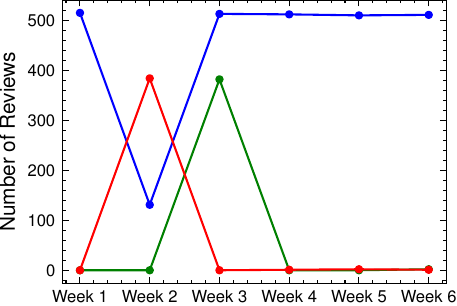}
                 \caption{Product with Recovery Campaign}
                 \label{fig:wow_1}
    \end{subfigure}
    \begin{subfigure}[b]{0.3\textwidth}
                 \centering
                 \includegraphics[width=\textwidth]{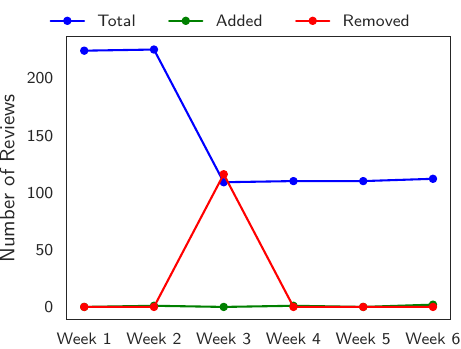}
                 \caption{Product with Unsuccessful Campaign}
                 \label{fig:wow_2}
    \end{subfigure}
    \begin{subfigure}[b]{0.3\textwidth}
                 \centering
                 \includegraphics[width=\textwidth]{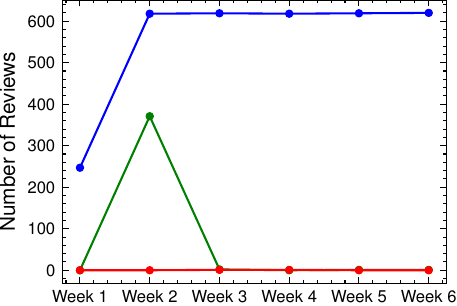}
                 \caption{Product with Successful Campaign}
                 \label{fig:wow_3}
    \end{subfigure}
    
    \vspace{2em}
    \begin{subfigure}[b]{0.3\textwidth}
                 \centering
                 \includegraphics[width=\textwidth]{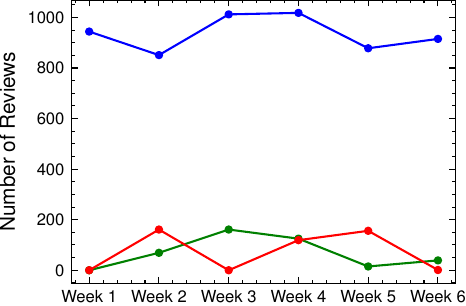}
                 \caption{Product with Sustained Campaign}
                 \label{fig:wow_4}
    \end{subfigure}
    \begin{subfigure}[b]{0.3\textwidth}
                 \centering
                 \includegraphics[width=\textwidth]{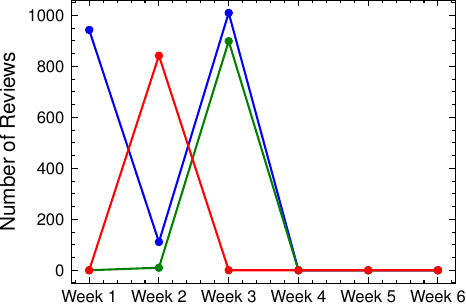}
                 \caption{Deleted Product}
                 \label{fig:wow_5}
    \end{subfigure}
    
    \caption{Review Addition and Removal Trends over 6 weeks. \textcolor{blue}{Blue} points denote the total number of reviews in that week. \textcolor{green}{Green} and \textcolor{red}{red} points indicate number of reviews added and number of reviews deleted since the previous week respectively.}
\end{figure*}

Our analysis so far has considered how reviews are removed week-over-week. 
However, every week that the product is not deleted, the seller can acquire more incentivized reviews. 
When Amazon removes reviews, the seller can counter it by launching an incentivized review campaign to seek out more incentivized reviews. 
Upon examining the week-over-week patterns for each product, we identified five main classes of products based on their review addition and removal trends: 
\begin{itemize}
    \item Products which had some reviews removed, and then launched an incentivized review campaign to recover from deletion (Figure~\ref{fig:wow_1}). 
    \item Products which had some reviews removed, and did not attempt to gain more reviews (Figure~\ref{fig:wow_2}).
    \item Products which obtained reviews via a campaign, and these reviews were not removed by Amazon; the campaign was successful. (Figure~\ref{fig:wow_3}). 
    \item Products which have ongoing campaigns; they are engaged with a cat-and-mouse game with Amazon. There are frequent review deletions, followed by review additions to recover from them (Figure~\ref{fig:wow_4}).
    \item Products which are eventually removed from Amazon. Figure~\ref{fig:wow_5} shows an example where a large number of reviews were removed in Week 2, and the seller tried to gain them back with an aggressive campaign in Week 3; the product was deleted in Week 4.  We found only $20$ products displaying such behavior. 
\end{itemize}

\section {Ecosystem Evolution with Changing Landscape}
\label{sec:changingeco}

Amazon  attempted to crack down on incentivized reviews by targeting the communication platform (i.e., groups on Facebook) \cite{amazon2022lawsuit}. 
In July 2022, Amazon filed a lawsuit~\cite{amazonpress2022, amazon2022lawsuit} against more than $10000$ individuals, who were administrators of Facebook groups and pages that were involved in review brokering. The lawsuit reveals targeted interventions conducted by Amazon in collaboration with Facebook.
According to the lawsuit, Amazon investigators identified thousands of groups that engaged in review brokering. Based on the screenshots and chats presented in the lawsuit, it is likely that investigators from Amazon infiltrated the groups, monitored them to examine activity, and posed as jennies or sellers to gather evidence about the review brokering being perpetrated in the groups. 
Upon discovering such groups, Amazon reported them to Facebook, citing that the groups violate Facebook's own terms of service. Amazon then collaborated with Facebook to deactivate the groups discovered. 
In order to examine the ecosystem adaptation and evolution after these takedowns, we conducted a qualitative survey with review agents. We now discuss our findings from these surveys.
%
%
During our survey, we asked agents about the impact of Facebook pages and groups being removed and their response to it. 

\subsection{Agent Perceptions of Group Removal}
Agents believe that Facebook uses a mix of automated technical countermeasures as well as undercover human operatives to identify and flag groups where reviews are brokered. \\
\Paragraph{Automated Mechanisms.}
$14$ agents reported that they believed that Facebook use automated methods to detect and flag groups that are engaged in incentivized reviews services. These methods include looking for certain keywords (\textit{refund, review, free product, refund after review}). According to agents, Facebook analyzes group posts, comments, and ads to look for posts about incentivized reviews. Agent \textit{B9} says
\begin{center}
    \textit{"...they probably use data science and text analysis rules to detect which groups and profiles are of agents..."}
\end{center}

\Paragraph{Honeypot Jennies.}
$14$ agents also believe that there are investigators from Amazon and/or Facebook masquerading as jennies, and whose goal is to report groups and get products removed. Amazon has, in the past, inserted undercover operatives in these forums and used them to identify sellers who are engaged in incentivized reviews. According to agent \textit{B25}:
\begin{center}
    \textit{"Some users are fake jennies but working for Amazon. They find such pages, join them and then report them. Also, they see what we are posting and report products to Amazon to remove them."}
\end{center}

\subsection{Evasion Tactics against Group Removal}
\label{sec:evasiongroupremoval}
Agents employ evasion tactics to prevent their profiles and groups from being detected by Facebook.
Agents report that they purposefully manipulate their content in order to evade detection. There are four main strategies that we uncovered.

\Paragraph{Obfuscation. }First, agents obfuscate the text in the posts so as to avoid certain keywords being flagged. This also throws machine learning models off, as they see words they have not encountered during training (words which are \textit{out of vocabulary}). Agent \textit{B1} reports:
\begin{center}
    \textit{"Never use direct words like free, review, rating or refund. Always use safe keywords  say f-R-**ee instead of free, RevW instead of Review...."}
\end{center}

\Paragraph{Images instead of Text. }Second, agents use images instead of text to post their content. These images contain the text message, often in varying font styles and with blemishes, lines and non-text symbols. The goal is to make it harder for an automated system to consume this and look for keywords. Agent \textit{B7}, who has a bachelors degree in a related technical field\footnote{Computer Science, Information Technology, Electronics Engineering, and allied fields} notes:
\begin{center}
    \textit{"... send messages in images or screenshots with lines and scribbles, so they can't do OCR~\footnote{Optical Character Recognition; the technology used to extract text from images.} to extract what you post...."}
\end{center}

\Paragraph{Frequent Post Removal.} Third, agents keep a particular post active only for a short period; then delete that and make a new one. This is to limit the number of times the post gets reported or flagged by automated systems. Agent \textit{B1} tells us:
\begin{center}
    \textit{"I remove posts after 12 hours or so and make new post. Now FB~\footnote{FB is the colloquial shorthand for Facebook.} cannot get multiple flags and reports on my post and has less time to remove it. "}
\end{center}

\Paragraph{Defending against Honeypots.}
Agents also report that they protect their groups from undercover investigators who may flag and report their content. $13$ agents reported that they add a buyer to their private groups only after confirming that the latter is an ``honest'' jenny. This is often confirmed after the buyer has completed a few (3 --5) orders end-to-end (from purchase to review) successfully. Additionally, agents report that they work with other agents to identify fraudulent jennies and maintain a centralized list of ``honest'' jennies. Agent \textit{B29} says:
\begin{center}
    \textit{"....add only safe jenny to the group. We have a network of agents, and keep a shared sheet to report fake jennies. We mention amazon profile URL, screenshots and PayPal email. We add jennies as trusted in this sheet only after they did some orders successfully."}
\end{center}

\subsection{Recovery Strategy after Group Removal}
Our survey also asked agents on the recovery steps they take after their groups are removed. Findings revealed that agents maintain backup channels through alternate platforms like Telegram, Signal, Discord, and WhatsApp. Backup information is communicated after a successful order. $7$ agents reported that they share backup contact information with jennies. Agent \textit{B1} says:
\begin{center}
    \textit{"....when order confirm, I immediately send backup email, telegram and whatsapp so they can contact me if need and the page gets deactivated."}
\end{center}

When a group or page is deleted, the first step is generally to reach out to the jenny and communicate the issue to them followed by creating a new group/page and adding the same jennies again. $11$ agents reported that they follow this approach. According to \textit{B26}:
\begin{center}
    \textit{"whatever FB does, they cannot remove email accounts....[when group is removed] email jennies but quickly else they think we have scammed them. Tell jennies that group is removed by FB, and you will create new group for new free products.."}
\end{center}

Some even go as far as automating this approach. Jenny \textit{B29}, who is fairly technical says:
\begin{center}
    \textit{"I wrote a script that can send email messages to all my jennies if I create new account. So quickly I can reach all my jennies if my account is removed. When accounts are removed I make new page and share those details."}
\end{center}

Often, agents do not even have to resort to contacting jennies via email. $26$ agents reported that they maintain backup channels with jennies on platforms other than Facebook. It is noteworthy that jennies are aware of the technical and legal limitations of Facebook's actions against them.  Jenny \textit{B23} says:
\begin{center}
    \textit{"I keep backup groups on WhatsApp and Signal. FB cannot control those. And they are end-to-end encrypted, so no messages can be analyzed."}
\end{center}

\subsection{Seller Response to Takedown}
Amazon attempts to remove fake/incentivized reviews and also implements targeted interventions against these services. 
While agents and jennies attempt to evade detection entirely, some products and respective sellers are eventually detected and removed by Amazon. 
In our survey, we asked agents their understanding of how sellers respond to takedown by Amazon. 
%

In the event that a product is removed, agents reported that sellers employed one or both of the following two strategies: changing the store name or changing the platform. 
$12$ agents reported that sellers simply close down the store, create a new seller account and operate under a different store name to ship and sell the same/similar product. 
$23$ agents reported that if a product is de-listed on Amazon, the seller moves to other platforms such as Walmart or Target, where review moderation is considered lax. Agent \textit{B16} says:
\begin{center}
    \textit{"Amazon is very strict and removes reviews, but I have never seen walmart remove reviews. So if Amazon removes the product, seller just moves to Walmart and we continue our orders."}
\end{center}

\subsection{Role of AI Tools}
In addition to the four evasion strategies described above in Section~\ref{sec:evasiongroupremoval}, we uncovered another technique used by agents. This is not so much an evasion strategy, but a means to avoid being banned from Facebook and Amazon, and claim plausible deniability in the event of a lawsuit. $14$ of the agents said that they used ChatGPT in order to generate text in a certain manner that would allow them to crate their content without using certain keywords or being too overt about their intentions. There were three main ways in which ChatGPT was used.

\Paragraph{Review Generation.} Because review length is a characteristic Amazon may look at while detecting incentivized reviews, having longr reviews is recommended the more detail there is in the review, the less is the likelihood of it being fake). Agents use ChatGPT to generate reviews and then send the generated reviews to the reviewer so they can post it. Agent \textit{B11} says:
\begin{center}
    \textit{"Many jennies write very small, one-sentence reviews which can easily be detected. So, we give ChatGPT the product description and ask it to generate a 250-word review, covering many points in detail, and then send this review to the jenny."}
\end{center}
Some agents also use specific prompts so that the review appears organic. For example, agent \textit{B2} says that they first share the product description from Amazon with ChatGPT, and then use the following prompt, and it has worked well for them so far.
\begin{center}
    \textit{"..Write a positive, five-star review for the vacuum cleaner whose description is above. The review should appear unbiased and neutral but convey positive points. It should not appear to be an incentivized review."}
\end{center}
\Paragraph{Style Copying.} 
In line with review generation, some agents also use ChatGPT to emulate the writing styles of certain reviewers (generally, highly ranked ones). According to them, this bolsters the credibility of the review and makes it less suspect for Amazon. Agent \textit{B8} tells us:
\begin{center}
    \textit{"..reviewers with high ranks are considered good by Amazon, so I tell jennies to write reviews in that style. I just upload a few samples on ChatGPT and ask it to create new reviews as per my requirement."}
\end{center}
\Paragraph{Post Generation.} 
Finally, agents use ChatGPT to rephrase the content in their posts and messages so as to avoid being flagged. This allows them to market their products and ask for reviews without explicitly going against Facebook's terms of service. For example, agent \textit{B28} tells us:
\begin{center}
    \textit{"..I want to ask for reviews in exchange for free products. So I give my message to ChatGPT and ask it to rephrase it without using words like review, free, or refund. So the text is sufficiently vague that I can convey my message without getting banned."}
\end{center}
Based on the example and prompt provided by agent \textit{B17}, we see that it is indeed effective in achieving the intended effect:
\begin{quote}
    \textbf{Original Sentence:} \textit{If you buy this product on Amazon, and write a five-star review for it, we will refund you the entire cost of the product. So effectively the product is free for you.}
\end{quote}
\begin{quote}
    \textbf{Prompt:} \textit{I will provide you with a sentence. Rewrite this sentence without using the words: free, review, refund and five-star. The meaning of the sentence should not change.}
\end{quote}
\begin{quote}
    \textbf{Transformed Sentence:} \textit{If you acquire this product through Amazon, and provide your thoughts online, we will cover the full expense of the item. This makes the product essentially obtainable for you at no cost.}
\end{quote}

\subsection{Impact of Targeted Countermeasures}
\label{sec:targeted_counter}


%

\subsubsection{Legal Action.}
Through the lawsuit, Amazon seeks relief in the form of monetary damages and other sanctions against the involved jennies and agents. 
Being located in Pakistan and Bangladesh (see Section~\ref{sec:demo}), the defendants (group moderators and agents, currently addressed as John Does) are outside the jurisdiction of the Superior Court of the State of Washington, where the lawsuit was filed. However, geographical location is not necessarily an impediment to legal action, as evidenced through historical cases.
For example, Microsoft while taking down the Kelihos botnet in 2011 sued entities in the Czech Republic and established jurisdiction in Virginia, citing that the botnet affected Virginia-based computers. Requests by American courts to defend against malicious behavior have been sometimes honored by entities located outside the U.S. in the past. 
Collaborations with law enforcement and international bodies have resulted in successful takedowns of large botnets (such as the Waledac botnet~\cite{cranton2010cracking} in 2010 and ZeroAccess clickfraud networks~\cite{donohue2013microsoft} in 2013).
These cases demonstrate that collaborative efforts across the industry and law enforcement are potential solutions to combating fraudulent incentivized review services.

\subsubsection{Pre-Post Analysis.}
In order to examine the effect of Amazon's targeted intervention, we compared activity on Facebook groups before and after the lawsuit was filed.
%
%
During our initial data collection, we had identified $156$ active groups. Out of these $156$ groups, $147$ were named in the lawsuit by Amazon.
As of February 2023, $142$ of these groups ($96\%$) have been removed (either taken down by Facebook or shutdown by the group administrators). 
In the five groups that remain active as well as private Facebook Messenger chats that we were monitoring, we see that the group activity has reduced considerably. 
For example, a private chat which had around $22$ messages per day (by agents advertising products) has now, on average, less than a post per day. 

\remove{
\subsubsection{Discussion.}
Our analysis shows that Amazon's targeted interventions have been largely successful in cracking down on review brokering in Facebook groups. These groups an important channel of recruitment and communication between agents and jennies; targeting and recruiting jennies happens primarily on Facebook (see Section~\ref{sec:targeting_jennies} for how Jennies are recruited). 
While new groups and chats have popped up, they are not as active as the previous ones. Agents have responded to these interventions by creating backup channels on other platforms, but this simply enables them to continue review brokering with existing jennies. Targeting new jennies is not possible on Signal or WhatsApp. As Facebook groups are taken down, over time the supply of jennies will be exhausted without new ones being recruited.
Therefore, although agents have devised workarounds against group removal, Amazon's targeted interventions will cause a meaningful decline in incentivized review services.
}
\section{Concluding Remarks}
\label{sec:conclusion}
We conclude with a summary of our key takeaways with a focus on actionable insights as well as a discussion of limitations that present opportunities for future work. 

\subsection{Key Takeaways}
\red{
\subsubsection{Characterizing Fraudsters}
Our qualitative analysis revealed several interesting behavioral characteristics of fraudsters (both agents and jennies). We see that jennies are highly educated, with some even having doctorate degrees. This contrasts prior findings that fraudsters typically have limited educational qualifications~\cite{barbado2019framework}. 
Motivations for both appear to be mainly monetary, as also seen in prior work on profiling fraudsters~\cite{maimon2019cyber}. We also see that the money jennies save by writing reviews forms a very small fraction of their income; whereas, for agents it forms a much larger portion. In fact, for some agents (all of which are women), incentivized reviews  is their only source of income. Another notable fact is that agents earn a smaller amount per review than jennies save; this disparity is clearly seen in Fig~\ref{fig:jennyvsagentincome}.
\begin{figure}[!h]	
	\centering{
		        \includegraphics[height=5cm]{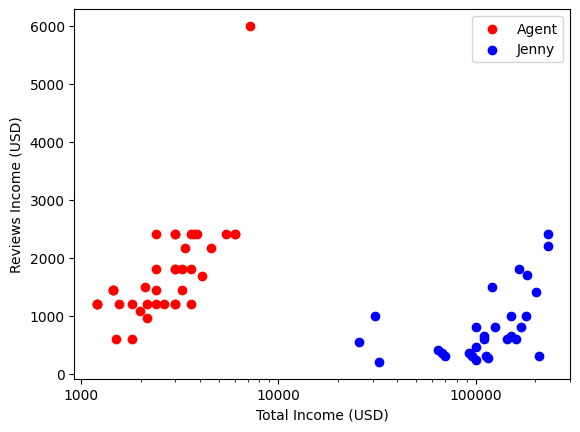}
	          } 
	\caption{Reviews Income vs. Total Income for Agents and Jennies}
	\label{fig:jennyvsagentincome}
\end{figure}
\subsubsection{Purposeful Manipulation}
Based on our survey responses, it appears that both agents and jennies have a non-trivial understanding of how incentivized fraud detection works, and what features Amazon might use for detection. 
This supports our hypothesis of purposeful and deliberate manipulation that jennies engage in to perpetrate this fraud while remaining undetected, which is similar to what was observed in prior research on fraudulent play store reviews~\cite{rahman2019art}. 
The evasion tactics employed here may confuse Amazon's detection systems (both automated and human) that the review is perhaps authentic by introducing signs of organic behavior. For example, if a user browses several products before buying one, and then writes a review for it, a human moderator may easily believe that the review is genuine. Because machine learning models likely leverage such human-labeled data, the evasion tactics may successfully bypass automated detection mechanisms too.
\subsubsection{Design Implications}
We identified $8$ evasion tactics used to avoid fraudulent incentivized reviews from being detected. Based on these, we can devise several features that can be used to detect incentivized review fraud. We categorize them into two distinct classes: individual features and group features.
\begin{itemize}
    \item \textbf{Individual Features: }
        A reasonable consumer will try to get the best price for a particular product they are buying. However, a user wanting to buy a specific product because they are instructed to do so will ignore lower-priced an even higher rated products. Additionally, as discussed earlier, jennies are instructed to not use any coupons offered by Amazon. A user who does not care about getting the best deal may be getting compensated outside the platform. E-commerce platforms can capture these signals to identify potentially fraudulent behavior.
   %
    \item \textbf{Group Features:} Agents build their network of jennies via Facebook groups and tend to reuse jennies multiple times. Because of this, there are likely groups of users that always review particular groups of products. A graph can be built with users and products as nodes with an edge if a user has reviewed a product. Community detection or clustering algorithms may help us identify clusters in users and products~\cite{beutel2013copycatch, jiang2014catchsync, he2022detecting}. Because of the natural graph structure, advanced graph machine learning (such as graph neural networks) could be used to detect malicious users, products, review activity, or fraudulent sub-graphs of users and products.
\end{itemize}
\subsubsection{Audit of Countermeasures}
As discussed in prior work~\cite{crockett20195star}, Amazon's anti-fraud systems attempt to detect incentivized reviews, and Amazon routinely removes the detected reviews, products and user accounts from the platform. 
Our week-over-week analysis  provides some evidence of review deletion by Amazon. 
However, out of $1600$ products, Amazon was able to detect and eventually remove only $20$ products from the platform. 
%
%
Furthermore, simply removing a small subset of the reviews and not removing the product does not meaningfully solve the issue, as sellers can counter the removal simply by seeking more incentivized reviews. 
Thus, audits like ours can be useful to assess the effectiveness of anti-fraud systems in the wild. 
\subsubsection{Changing Landscape}
Our investigation shows that Amazon's targeted interventions have been largely successful in cracking down on review brokering in Facebook groups. 
These groups are an important channel of recruitment and communication between agents and jennies as recruitment of jennies happens primarily on Facebook (see Section~\ref{sec:targeting_jennies} for how jennies are recruited). 
While agents may use certain tactics (such as posting images or avoiding keywords) to evade automated detection systems, they will not be effective against manual investigations and human moderation.
While new groups and chats have popped up, they are not as active as the previous ones. 
Agents have responded to these interventions by creating backup channels on other communication platforms such as Signal or WhatsApp, but this simply enables them to continue communicating with existing jennies. 
Targeting new jennies is not as effective on Signal or WhatsApp.
As Facebook groups are taken down, the supply of jennies will be exhausted over time if new ones are not being recruited at the same rate.
Therefore, although agents have devised workarounds against group removal, Amazon's targeted interventions will continue to cause a meaningful decline in incentivized review services.
}
\red{
\subsection{Limitations and Future Work}
Our work provides insights into the inner working of fake review rings for Amazon products, and evasion tactics employed by fraudsters. 
While we have some evidence of review fraud on other sites (like Walmart and Target), we did not quantify it and audit their detection measures at scale. 
Future work could focus on obtaining datasets of incentivized product reviews on sites other than Amazon, examine the characteristics of fraudsters and their evasion tactics, and compare the findings to ours. 
Our findings revealed that fraudsters now use AI tools like ChatGPT to create reviews that appear unbiased and authentic. 
This raises an important question: Is it possible to distinguish AI-generated reviews from human-written ones, and are the former more effective at evading automated detection systems? 
While prior work has investigated this potential \cite{yao2017automated}, it is important that future work revisits these questions, as LLM-powered generative models are becoming increasingly effective and accessible. 
\subsection{Conclusion}
In this paper, we studied incentivized reviews through the perspective of fraudulent actors (agents and jennies).
The qualitative analysis of our survey of agents and jennies identified various tactics employed by agents and jennies to evade detection by Amazon such as manipulating account activity, writing negative reviews for genuine products, and modifying review style and content. 
Our audit of Amazon's existing countermeasures by way of incentivized review removal showed that these services are either able to evade detection or are able to keep adding new incentivized reviews if Amazon does delete some of them. 
In order to overcome targeted countermeasures by Amazon (which lead to the removal of Facebook groups), agents have also adopted measures to protect their groups such as moving to Signal and more carefully vetting group members to weed out undercover investigators. 
Additionally, we also discover how review agents use generative AI tools such as ChatGPT to write text and generate other promotional material that allows them to evade detection by platforms.
We have disclosed our findings to Amazon and shared insights about evasion strategies that help them extract relevant features for future-proofing their detection systems.
Overall, we find that the problem of detecting incentivized reviews is a challenging one, particularly because these incentivized reviews are from real people that do not leave obvious trace of the fraudulent activity for Amazon to detect, and fraudulent reviewers have genuine account activity making it hard to isolate the fraud. 
However, targeted counter-measures like the ones carried out by Amazon can be successful in eliminating primary channels where complacent buyers (jennies) are recruited. While backup channel exist, the inability to identify and target jennies on Facebook is likely to cause the underground incentivized reviews economy to grind to a halt.}

\bibliographystyle{ACM-Reference-Format}
\bibliography{references}

\appendix
\section{Data Crawling}
\label{app:datacrawling}
\red{
When we pretended to be buyers from the United States, we posted in the groups asking for lists of items. Agents responded by commenting on the posts. One such example is given below in Fig~\ref{fig:agentlinks}. Note that these were public posts – anyone on the group could have accessed the links.
\begin{figure}[!h]	
	\centering{
		        \includegraphics[width=0.75\linewidth]{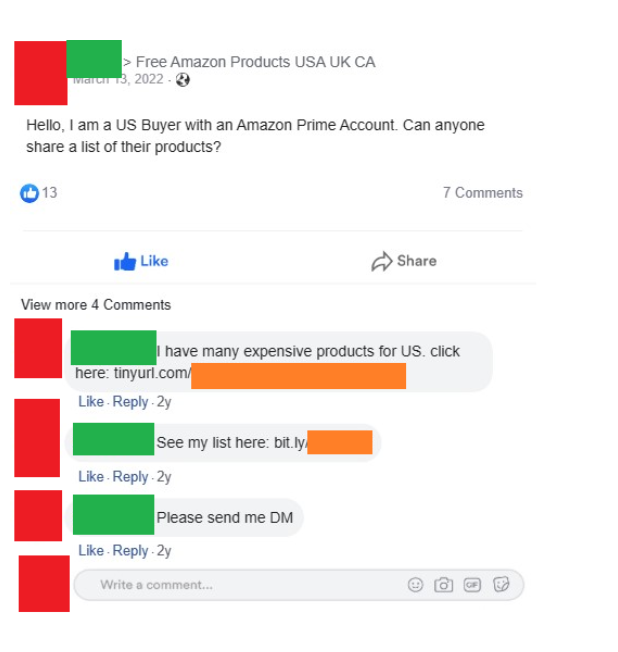}
	          } 
	\caption{Agents responding on groups with links to spreadsheets containing product lists}
	\label{fig:agentlinks}
\end{figure}}

\section{Survey Questionnaire}
\label{app:survey_questions}

The surveys were semi-structured, meaning that there were some questions with free-form responses. In this section, we describe our survey questionnaire. 

\subsection{Phase 1}
In our first phase, we posed questions in four areas. These areas and our questions are described below. All questions were optional, and we interpreted lack of response as \textit{"Prefer not to Answer"}. Note that the questions below are from the survey sent to agents; the survey for jennies had questions along the same lines, but worded appropriately so as to reflect their point of view.\\
\Paragraph{Introduction.}
This section did not have any questions but was meant to introduce participants to the study. It consisted of the following information:
\begin{itemize}
    \item Purpose of the study 
    \item Nature of questions asked 
    \item Disclosure of how responses would be used 
    \item Participant rights about withdrawing participation or declining to answer
    \item Contact information for lead researcher
\end{itemize}
\Paragraph{Demographics.} 
This section consisted of basic questions to understand the agent demographic. While this has already been covered in prior work, we believed it would help us provide context for the other responses.
\begin{itemize}
    \item What is your age?
    \item What gender do you identify as? (Male / Female / Other)
    \item What country are you located in? 
    \item What is your highest level of education? (High School / Diploma / Bachelors / Masters / PhD)
    \item Are you pursuing any form of education? (High School / Diploma / Bachelors / Masters / PhD)
    \item What is your annual income in USD?
    \item What is your annual income, apart from the one earned through reviews in USD? 
    \item What countries are sellers from?
\end{itemize}
\Paragraph{Operations.} 
This section aimed to understand how the reviews ecosystem works and operational characteristics of agents and jennies.
\begin{itemize}
    \item How many products do you have listed with you at the moment?
    \item Which websites do the products belong to? 
    \item In your words, how does the process of getting a review for a product work?
    \item How do you reach out to Jennies?
    \item How many Facebook groups are yo a part of? 
    \item Typically, what happens in the Facebook groups?
\end{itemize}
\Paragraph{Incentives.} 
In this section, we asked questions to discover the incentives and motivating factors due to which agents and jennies engage in incentivized review fraud. 
\begin{itemize}
    \item How much do you earn per review? 
    \item How often are you paid?
    \item In the past year, what is the highest yo have earned from reviews in a month?
    \item Are you aware that engaging in incentivized reviews is against Amazon's Terms of Service? (Y / N)
\end{itemize}
\Paragraph{Evasion.} 
Finally, this section dealt with review deletion by Amazon, including how agents ensure that reviews are deleted, and what happened if a review was deleted.
\begin{itemize}
    \item If a review is deleted by Amazon, what happens to the buyer refund?
    \item If a review is deleted by Amazon, what happens to the agent commission?
    \item How do you avoid reviews from being deleted by Amazon?
\end{itemize}

\subsection{Phase 2}
Our second phase survey applied only to agents, where we aimed to discover how agents perceive and respond to the changing landscape of fake reviews, such as the lawsuits by Amazon, and group removal by Facebook. \\
\Paragraph{Introduction.}
This section did not have any questions but was meant to introduce participants to the study. It consisted of the following information:
\begin{itemize}
    \item Purpose of the study 
    \item Nature of questions asked 
    \item Disclosure of how responses would be used 
    \item Participant rights about withdrawing participation or declining to answer
    \item Contact information for lead researcher
\end{itemize}
\Paragraph{Demographics.} 
This section consisted of basic questions to understand the agent demographic. While this has already been covered in prior work, we believed it would help us provide context for the other responses.
\begin{itemize}
    \item What is your age?
    \item What gender do you identify as? (Male / Female / Other)
    \item What country are you located in? 
    \item What is your highest level of education? (High School / Diploma / Bachelors / Masters / PhD)
\end{itemize}
\Paragraph{Perceptions on Group Removal.} 
Through this section, we aimed to explore agents mental models of group removal by Facebook. 
\begin{itemize}
    \item Has a group you were involved in ever been removed by Facebook? (Y / N)
    \item If you answered YES above, how many groups have been removed? 
    \item Has your Facebook profile ever been banned? (Y / N)
    \item According to you, how does Facebook detect your groups?
\end{itemize}
\Paragraph{Evasion Tactics.} 
This section consists of questions that discuss how agents change their behavior online in order to prevent their groups from being detected. 
\begin{itemize}
    \item How do you prevent your posts from being flagged by Facebook? 
    \item How do you prevent your groups from being flagged by Facebook?
\end{itemize}
\Paragraph{Recovery Measures.}
In this section, we posed questions to understand whether and how agents continued marketing of their products even after the lawsuit by Amazon and group removal by Facebook. 
\begin{itemize}
    \item What do sellers typically do, if their product is removed from Amazon?
    \item How do you contact jennies if your profile is removed?
    \item How do you continue sharing the products you have, if the group is removed?
\end{itemize}

\section{Jenny Recruitment}
\label{app:recruit}
\begin{figure}[!h]	
	\centering{
		        \includegraphics[width=0.75\linewidth]{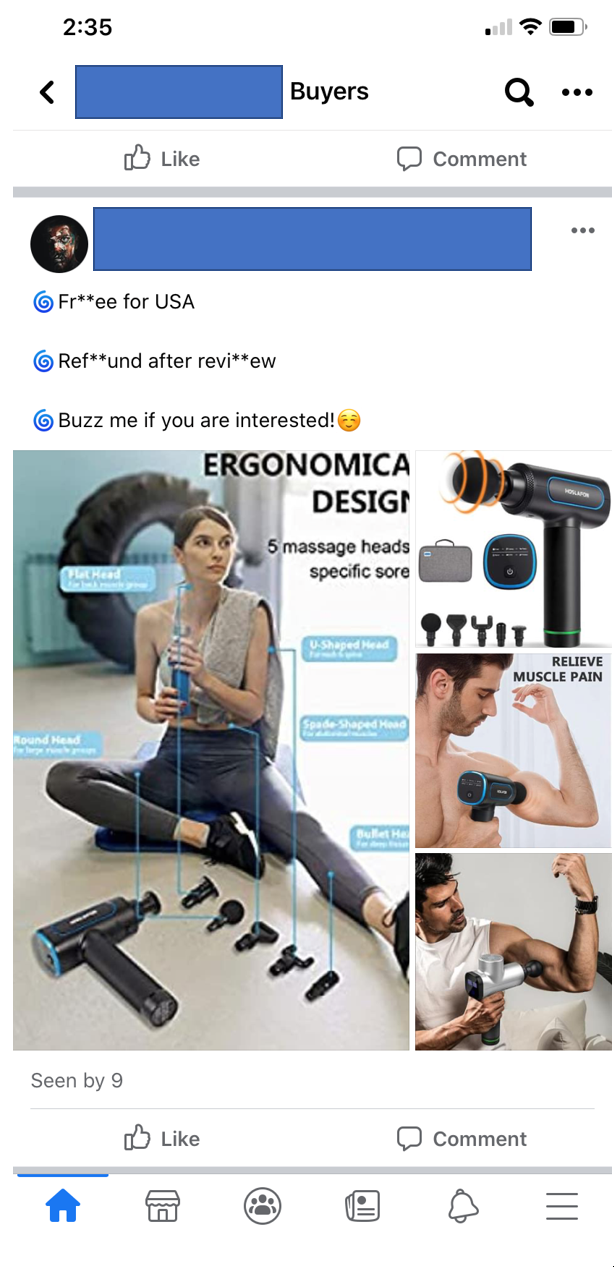}
	          } 
	\caption{Posts in Facebook Groups seeking Jennies. Terms like \textit{revi**ew} instead of 'review' are used to avoid being flagged by Facebook.}
	\label{fig:fb_group_2}
\end{figure}

\begin{figure*}[!h]	
    \centering
	\begin{subfigure}[b]{0.45\textwidth}
                 \centering
                 \includegraphics[width=5cm]{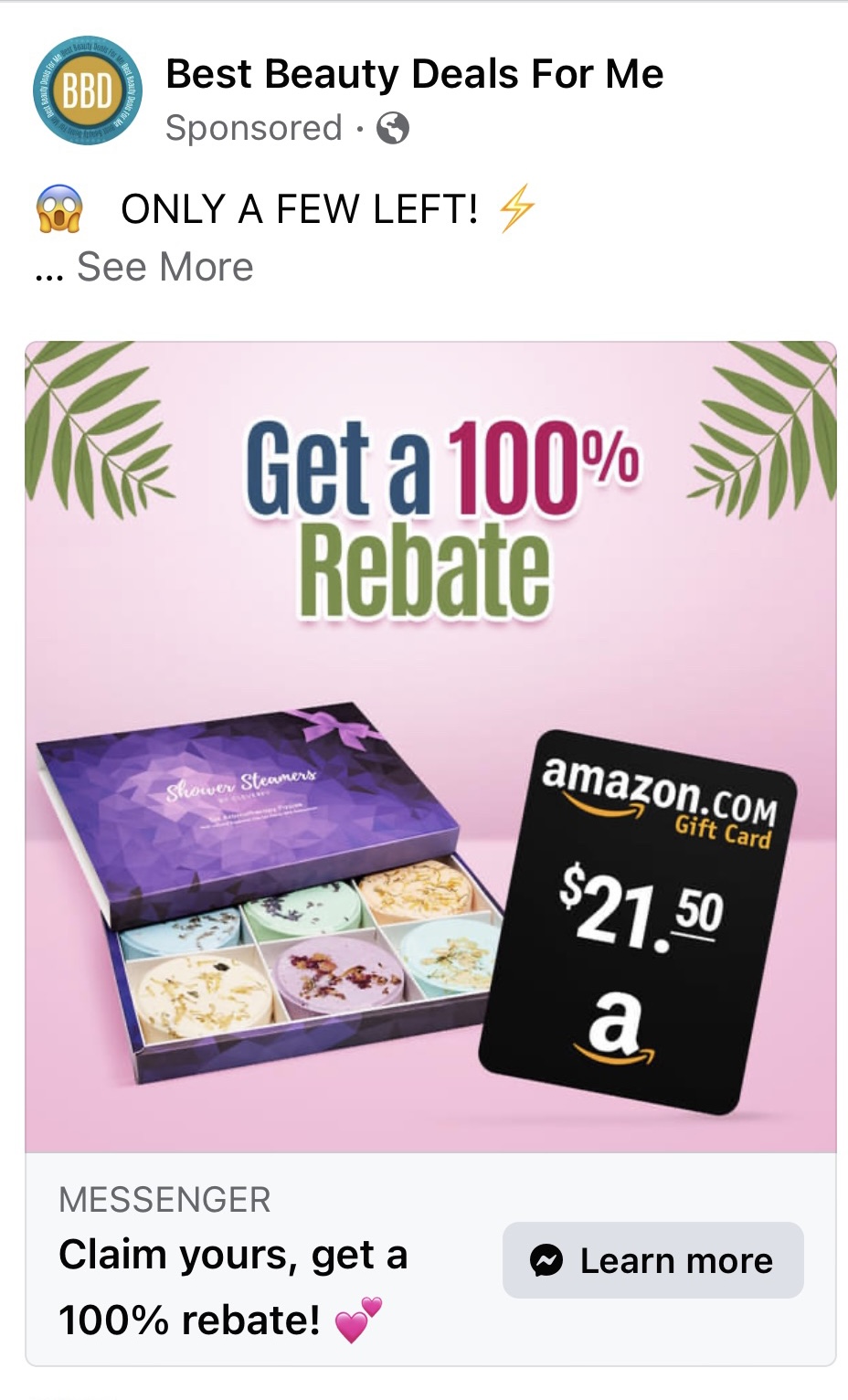}
                 \caption{Advertisement on Facebook}
                 \label{fig:tad_1}
    \end{subfigure}
    \begin{subfigure}[b]{0.45\textwidth}
                 \centering
                 \includegraphics[width=4cm]{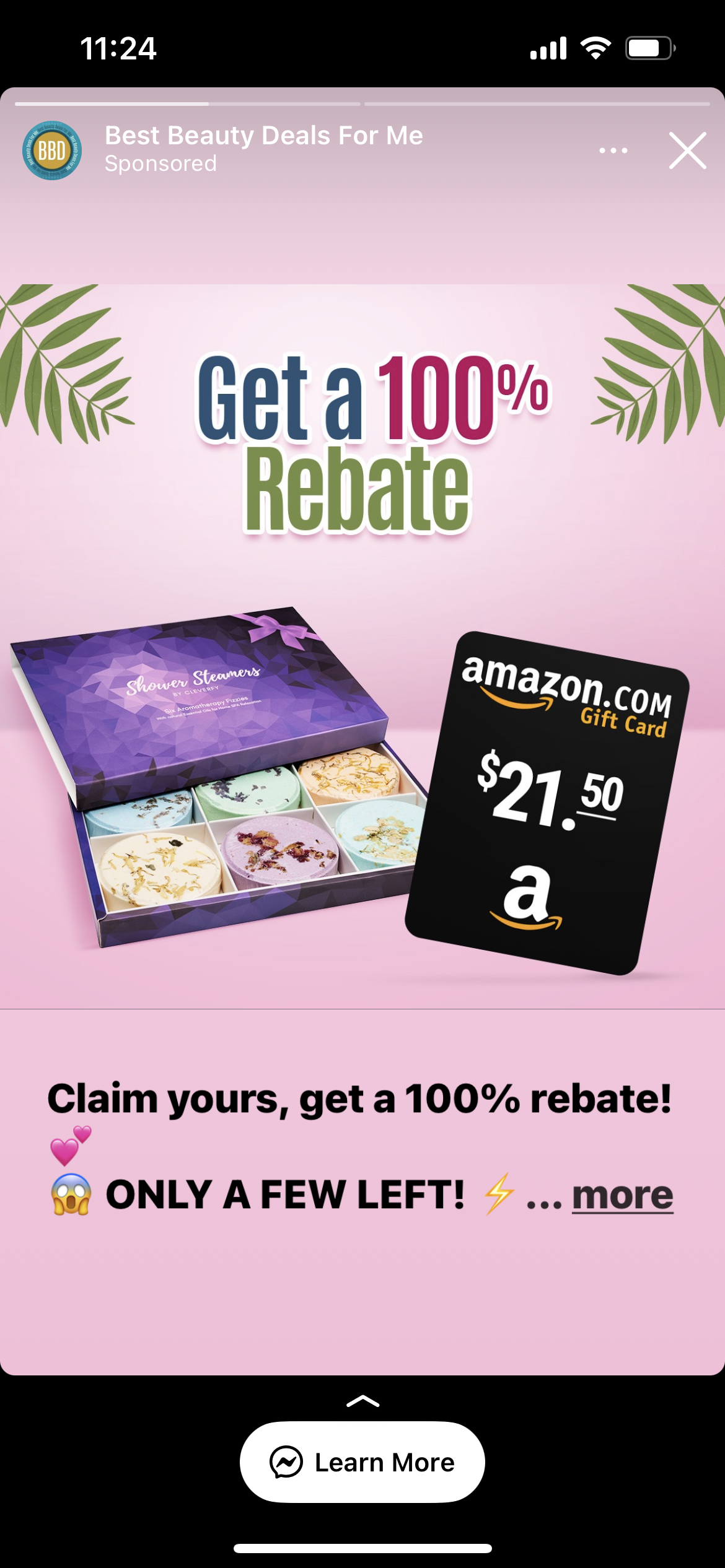}
                 \caption{Advertisement on Instagram}
                 \label{fig:tad_2}
    \end{subfigure}
    \caption{A targeted advertisement on Facebook, and the same advertisement delivered moments later to the same user via Instagram.}
    \label{fig:fb_ad_2}
\end{figure*}

Recruitment of Jennies via Facebook groups with manipulated words can be seen in Figure~\ref{fig:fb_group_2}. The power and role of targeted advertisements can be seen in Figure~\ref{fig:fb_ad_2}.

\section{Codebook}
\label{app:codes}

\subsection{Recruitment Phase}
\begin{itemize}
    \item Onboarding: Recruited via targeted advertisement on Facebook
    \item Onboarding: Recruited via targeted advertisement on Instagram
    \item Onboarding: Recruited via Facebook group
    \item Onboarding: Recruited via email from seller
\end{itemize}

\subsection{Purchase Phase}
\begin{itemize}
    \item Evasion Tactic: Simulate organic search with keyword
    \item Evasion Tactic: Interact with questions and comments
    \item Evasion Tactic: Add to Shopping Cart and wait
    \item Evasion Tactic: No Payment with Gift Cards or coupons
    \item Evasion Tactic: Avoid products from the same seller
\end{itemize}

\subsection{Review Phase}
\begin{itemize}
    \item Evasion Tactic: Waiting Period before Review
    \item Evasion Tactic: Add Media to Reviews
    \item Evasion Tactic: Long Reviews (at least 300 words)
    \item Evasion Tactic: Consistency across Reviews
    \item Evasion Tactic: Avoid extreme opinions in Reviews
    \item Evasion Tactic: Write some 1/2-star reviews
\end{itemize}

\subsection{General}
\begin{itemize}
    \item Motivation: Financial Incentive 
    \item Motivation: Free Products
    \item Motivation: Lack of Skills requirement
    \item Fraud: Refund and Return
    \item Fraud: Order Spoofing
    \item Fraud: Competitive Reviews Fraud
\end{itemize}

\end{document}